\documentclass[11pt,a4paper]{article}
\usepackage[hyperref]{acl2022}
\usepackage{times}
\usepackage[utf8]{vietnam}
\usepackage{latexsym}

\usepackage{graphicx}
\graphicspath{{./images/}}
\usepackage{microtype}

\aclfinalcopy 

\title{An Empirical Study on Learning Latent Representations for Emotional Speech Synthesis}

\author{Dang Quang Vinh \\
  Aimesoft JSC \\
  Hanoi, Vietnam \\
  \texttt{vinhdq@aimesoft.com} \\\And
  Ngo Quang Huy \\
  Aimesoft JSC \\
  Hanoi, Vietnam \\
  \texttt{huynq@aimesoft.com} \\}

\date{}

\begin{document}
\maketitle
\begin{abstract}
For the last couple of years, the field of speech synthesis has improved dramatically thanks to deep learning. There are more and more deep learning-based TTS systems developed to make it possible to produce voices with high intelligibility and naturalness. Meanwhile, controlling the expressiveness is yet a big deal, generating speech in different styles or manners has received a lot of attention from community recently. This paper aims to give our solutions to deal with the task emotional speech synthesis (ESS) at VLSP 2022 which allows to generate humanlike natural-sounding voice from a given input text with desired emotional expression. By integrating speaker embedding, prosody bottleneck into FastSpeech 2, our systems can promisingly generate emotional speech of a single speaker (Sub-task 1), transfer speaking styles from another speaker to the target speaker with neutral non-expressive data while retaining the target speaker's identity (Sub-task 2).
\end{abstract}

\section{Introduction}
\label{sec:intro}

Speech synthesis, also known as Text to speech (TTS), which aims to synthesize intelligible and natural speech from text \cite{taylor2009text}, plays a vital role in speech interaction systems, and can be widely applied in human communication \cite{adler2006understanding}. With the rise of deep learning, TTS has thrived exponentially and become more humanlike, current state of the art synthesizers still struggle to create speech as natural as possible but emotion simulation is not a self-evident feature. Lately, a very large amount of research work has come out focusing on every aspect of speech synthesis. Little work has been conducted that utilizes latent spaces to model the variability in speech of different styles or different emotions. An issue for such a control perhaps lies in the complexity of human vocal expression. For more than a decade now, tons of attempts to add emotion effects to synthesized speech have been made, which are based on old synthesis methods namely formant synthesis \citep{burkhardt2000verification, burkhardt2001simulation}, diphone concatenation \citep{vroomen1993duration, edgington1997investigating, heuft1996emotions}, unit selection \citep{iida2000speech, marumoto2000control}.

To promote research in emotional speech synthesis and related areas, VLSP 2022, the International Workshop on Vietnamese Language and Speech Processing, devotes task Emotional Speech Synthesis. The aim of the task is to build TTS systems that generate expressive speech. The task consists of 2 sub-tasks: synthesizing emotional speech of a single speaker given emotional training dataset, synthesizing emotional speech of another speaker, whose training dataset is neutral non-expressive data, provided dataset from the first sub-task (speaker adaptation). 

Numerous methods have been developed to deal with the emotional speech synthesis (both single speaker and and speaker adaptation). Most of them leverage latent embedding spaces integrated into end-to-end neural TTS frameworks that can be Global Style Tokens combined with Tacotron in the early days of neural TTS \cite{wang2018style} or recent variants of learning embedding space of prosody \cite{pan2021cross} combined with frameworks such as Tacotron 1/2, FastSpeech 2 \citep{skerry2018towards, lee2019robust, KlimkovRRD19}. The prosody bottleneck builds up the kernels accounting for speaking well, and disentangles the prosody from content and speaker timbre, therefore guarantees high quality speaker adaptation.

We empirically verify the effectiveness of the proposed method on the training dataset of VLSP 2022 Task Emotional Speech Synthesis. In spite of late evaluation from the organizers, synthesized audios show that our method is totally potential when they satisfy requirements of affective expression: understandable, natural-sounding and clearly expressive. 

\section{Proposed Methods}
\label{sec:methods}

\subsection{Data Preprocessing}

After receiving two datasets VLSP-EMO and VLSP-NEU from the organizers, we performed some tests and found quite many problems from them, especially VLSP-EMO.

As the dataset VLSP-EMO was collected from films and interviews, it contains a lot of noise, background music, and sometimes voices from other people or even "different voices" of a single speaker due to recording conditions (e.g. film versus interview). We employ Facebook Denoiser \cite{defossez2020real}, an encoder-decoder based architecture with skip-connections that can remove various kinds of background noise and room reverb, to first enhance the audio quality.

Besides, the provided text transcripts include some English words (e.g. me too, resort,..) and typos. We fixed the typos and splitted the English words into similarly-pronounced syllables in Vietnamese (e.g. mi tu, ri sọt,..).

Another problem is that there was a portion of audio files that were incorrectly transcribed. To resolve this, we used an Automatic Speech Recognition (ASR) tool to transcribe the audio files then compared the results with ground-truth transcripts using Character Error Rate (CER), filtered out and corrected the text labels with high CER.

Finally, we resampled all the audio files to the rate of 22050 Hz and removed ones that we considered unintelligible.


\subsection{Model Architecture}
In both sub-tasks, we used FastSpeech 2 \cite{ren2020fastspeech} with some modifications as the acoustic model. FastSpeech 2 was proved to generate speech significantly faster than previous auto-regressive models with comparable quality. It also provides further controllability on pitch, energy and duration, which are the key factors in expressive speech synthesis.

Our model takes as input a sequence of phones (phonemes) and produces spectrogram in log-mel scale. Similar to \cite{ren2020fastspeech}, we also used Montreal Forced Aligner \cite{mcauliffe17_interspeech} tool to extract the
phoneme duration for training.

We utilized a pre-trained HiFi-GAN \cite{kong2020hifi} V1 variation as the vocoder to synthesize final waveform from Mel-spectrogram.

\subsubsection{Sub-task 1 – ESS with a single speaker}
In sub-task 1, based on FastSpeech 2 we simply added a kind of emotion embedding - a look-up table that maps from an emotion id to a fixed-length vector.

The emotion vector is broadcast-added with the output of the encoder. The result then goes through the rest of the architecture and ends up as emotional Mel-spectrogram.

\begin{figure}
    \centering
    \includegraphics[scale=.25]{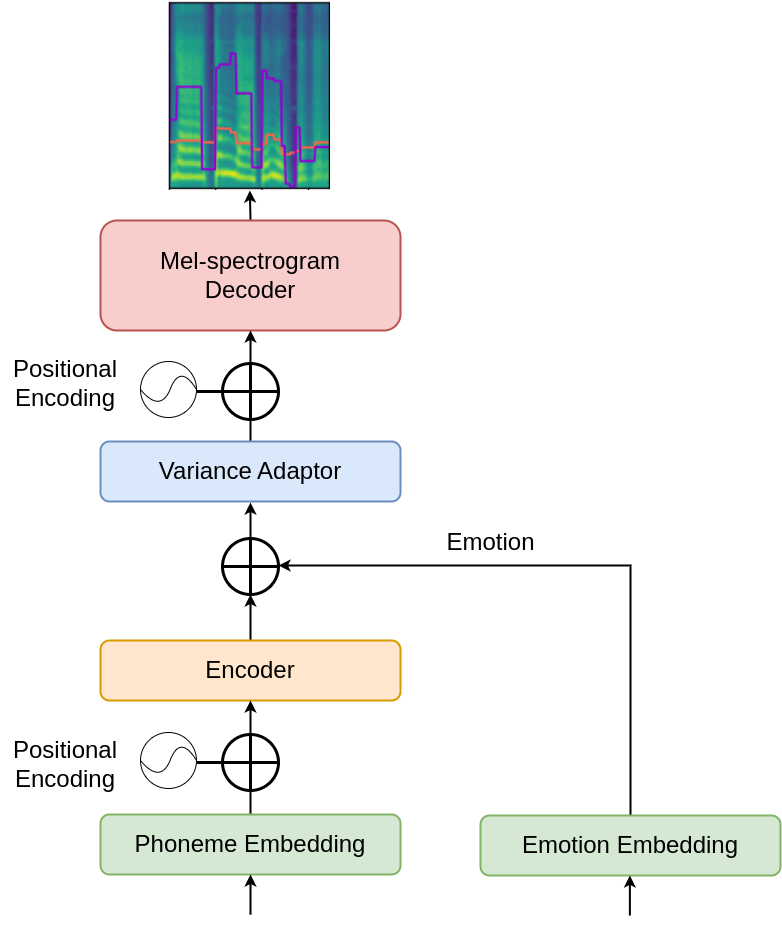}
    \caption{The architecture for sub-task 1}
    \label{fig:subtask1}
\end{figure}

The proposed structure is illustrated in Figure \ref{fig:subtask1}.

\subsubsection{Sub-task 2 – ESS with speaker adaptation}
As inspired by \cite{pan2021cross}, we expanded FastSpeech 2 with speaker and emotion embeddings, both of which are just look-up tables as used in sub-task 1.

Following \cite{pan2021cross}, the speaker embedding is first
broadcast-concatenated with the output of 
the encoder, the result of which is then projected back to the original embedding size. The emotion embedding is passed through a linear layer with tanh activation to a vector of the same dimension as encoder output, then added to the previous speaker-combined encoder output. The encoder output containing speaker and emotion information is referred to as speaker-emotion-combined encoder output. This encoder output afterward flows through a "prosody bottleneck" as in \cite{pan2021cross}, added back to the previous speaker-emotion-combined output via residual connection. The result of these above operations is hereafter fed to the Variant adaptor, then Decoder to finally generate target Mel-spectrogram.

We believed that the use of prosody bottleneck would retain the emotional information which helps in transferring emotion from source speaker to target speaker without losing the target speaker's timbre.

\begin{figure}[h]
    \centering
    \includegraphics[scale=.18]{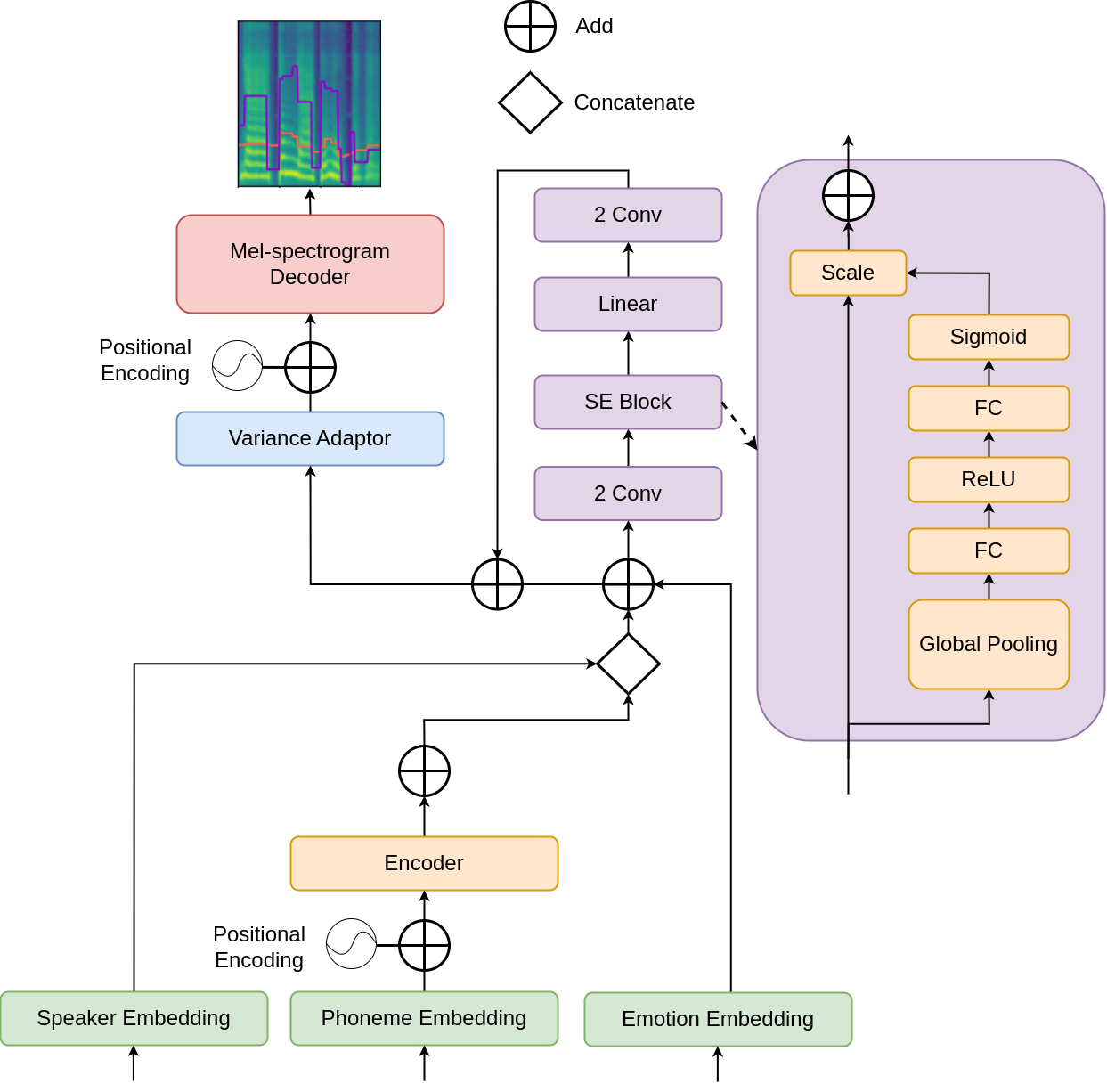}
    \caption{The architecture for sub-task 2}
    \label{fig:subtask2}
\end{figure}

The proposed structure is illustrated in Figure \ref{fig:subtask2}.

\section{Experiments and Results}
\label{sec:exp}

\subsection{Datasets}
Conforming to the organizer's regulations, we used only two provided datasets, VLSP-EMO and VLSP-NEU. After the preprocessing, we achieved 3.8 and 11.89 hours of speech audio of VLSP-EMO and VLSP-NEU, respectively. The use of these two datasets was as follows:

\begin{itemize}
    \item Sub-task 1: we used all data from VLSP-EMO for training.
    \item Sub-task 2: we combined VLSP-EMO and VLSP-NEU to form a large dataset and used it for training.
\end{itemize}

\subsection{Training Strategy}
For both sub-tasks we used the same training configuration. The modified FastSpeech 2 was trained from scratch for 40000 steps with batch size of 16. The used optimizer was Adam \cite{kingma2014adam} with $\beta_1 = 0.9$, $\beta_2 = 0.98$, $\epsilon = 10^{-9}$. We started training with 3000 warm-up steps then annealed the learning rate after milestone steps of 5000, 9000, 17000 with the rate of 0.3.

It took about an hour and more than 3 hours of training time on a GPU NVIDIA GeForce RTX 2080 Ti to create usable models for the two sub-tasks respectively.
\subsection{Experimental Results}
Our systems achieved the following results according to the organizers’ announcement:

\subsubsection*{Sub-task 1}
\begin{itemize}
    \item Naturalness Test: MOS = 2.719 / 5.
    \item Intelligibility Test (SUS): Syllable Error Rate = 72.40\%.
\end{itemize}

\subsubsection*{Sub-task 2}
\begin{itemize}
    \item Naturalness Test: MOS = 1.622 / 5.
    \item Intelligibility Test (SUS): Syllable Error Rate = 64.80\%. 
    \item Speaker Similarity: Similarity Score = 1.543 / 4.
\end{itemize}

\section{Conclusion}
\label{sec:conclusion}
We have presented our empirical study of learning latent representations for emotional speech synthesis task at VLSP 2022. By applying various latent embedding spaces, synthesized audios show that our method is a favourable approach to the benchmark dataset of VLSP 2022 TTS. Without using any external source, our method is straightforward to adapt to other languages. We plan to investigate latent space representations for other speech synthesis tasks, including multilingual speech synthesis and adaptive speech synthesis, among others.  
\bibliography{vlsp2022}
\bibliographystyle{acl_natbib}

\end{document}